\documentclass[sigconf, nonacm]{acmart}

\usepackage{soul}
\usepackage{xspace}

\AtBeginDocument{%
  }

\newif\ifCOMMENTS
\COMMENTStrue 

\ifCOMMENTS

\fi

\newcommand{\sophia}{\textbf{\emph{Sophia}}\xspace}

\usepackage{xcolor}

\newif\ifshowchanges
\showchangesfalse

\ifshowchanges
  \newcommand{\rev}[1]{\textcolor{blue}{#1}}
\else
  \newcommand{\rev}[1]{#1}
\fi

\begin{document}

\title{Foreign Domestic Workers' Perspectives on an LLM-Based Emotional Support tool for Caregiving Burden}


\author{Shin Shoon Nicholas Teng}
\email{nicholas_teng@mymail.sutd.edu.sg}
\orcid{0009-0002-8028-5589}
\affiliation{
  \institution{Singapore University of Technology and Design}
  \city{Singapore}
  \country{Singapore}
}

\author{Kenny Tsu Wei Choo}
\email{kenny_choo@sutd.edu.sg}
\orcid{0000-0003-3845-9143}
\affiliation{
  \institution{Singapore University of Technology and Design}
  \city{Singapore}
  \country{Singapore}
}


\begin{abstract}
Foreign Domestic Workers (FDWs) play a central role in home-based eldercare yet often experience substantial emotional caregiving burden shaped by linguistic barriers, social isolation, and limited access to support.
While caregiving burden has been extensively studied among familial caregivers, little is known about how FDWs engage with emotional support technologies.
We present an exploratory qualitative study of how FDWs in Singapore interact with a Large Language Model (LLM)–driven chatbot as an everyday, non-clinical form of emotional support. 
Through interviews and guided chatbot interactions, we conducted an inductive thematic analysis of participants’ experiences.
We identify three design-relevant themes: chatbots were experienced as psychologically safe and emotionally validating; they supported linguistic accessibility by accommodating imperfect and fragmented language; and they were appropriated as multifunctional resources for reassurance, guidance, and companionship.
We discuss implications for designing LLM-driven emotional support tools that foreground psychological safety, accessibility, and flexible appropriation.

\end{abstract}




\maketitle

\section{Introduction}
Foreign Domestic Workers (FDWs) play a central role in supporting home-based eldercare in ageing societies such as Singapore \cite{Devasahayam22022010, Yeoh01052010}. In many households, FDWs provide day-to-day assistance for frail older adults, managing both the practical and emotional demands of caregiving  \cite{TAM20181269}. Despite this central role, FDWs often experience substantial emotional caregiving burden shaped by linguistic barriers, social isolation, and structural constraints associated with migrant domestic work \cite{Yeoh01052010, Ha2018FDWBurdenSingapore}. These conditions can restrict opportunities for emotional expression, limit access to social support, and intensify feelings of stress, uncertainty, and loss of control in everyday caregiving situations.

While caregiving burden has traditionally been examined through the lens of familial caregivers such as spouses or adult children \cite {Lawton1989CaregivingAppraisal, Zarit1980Burden, Pearlin1990StressProcess}, this body of work implicitly marginalizes non-familial caregivers like FDWs, whose caregiving labor is situated within employment relationships rather than kinship ties. As a result, little is known about how FDWs emotionally appraise caregiving stress or seek support in everyday caregiving contexts where formal resources are limited or inaccessible.

Recent advances in Large Language Model (LLM)-driven chatbots present a potential form of everyday, non-clinical emotional support that may align with FDWs’ constraints. Chatbots offer private, low-cost, readily accessible interaction and have demonstrated promise in providing emotional support and fostering a sense of being heard through empathetic and reflective dialogue \cite{woebott2017, wysa2018}. For FDWs, who may face stigma, social isolation, and linguistic barriers when seeking help \cite{Ho2022PeerSupportMDW, Shrestha2019WorkerMobilitySingapore}, LLM-driven chatbots may offer a distinct form of emotional support that is difficult to access through existing social or institutional channels.

While prior research has examined the use of chatbots for mental health and caregiving support more broadly, there is limited understanding of how FDWs themselves perceive and engage with LLM-driven chatbots, and how these experiences can inform the design of emotional support technologies. In particular, it remains unclear how FDWs experience these systems as conversational partners, how linguistic and cultural factors shape interaction, and what expectations FDWs form regarding the role such tools might play in their caregiving lives.

To address this gap, we conducted an exploratory qualitative study with seven FDWs working in Singapore to examine how they interact with an LLM-driven chatbot when reflecting on emotionally challenging caregiving experiences. Specifically, this study seeks to understand how FDWs interact with LLM-driven chatbots in relation to managing their emotional caregiving burden. Therefore, we investigated the following research question: 

\begin{quote}
    "How might FDWs in Singapore interact with LLM-driven chatbots in relation to managing their caregiving burden?"
\end{quote}
By foregrounding FDWs’ perspectives, this study contributes qualitative insights into how FDWs experience LLM-driven emotional support in everyday contexts, and surfaces design implications for such systems that aim to support an FDW's emotional well-being.

\section{Related Work}
\subsection{Caregiving Burden and Emotional Appraisal}
Caregiving burden has been widely conceptualized as a subjective and multidimensional experience shaped not only by the objective demands of care, but by caregivers’ emotional appraisal of their caregiving role.
Pearlin's stress process model~\cite{Pearlin1990StressProcess} frames caregiving burden as arising from the interaction between caregiving-related stressors and the availability of social and emotional support.
In this view, caregiving burden intensifies not only as stressors accumulate, but also when caregivers lack access to emotional, social, or communicative forms of support that enable them to cope with and make sense of these demands.

For FDWs, these dynamics are often amplified by structural and communicative constraints associated with migrant domestic work.
Long working hours, limited rest days~\cite{Parreñas16122021}, and caregiving in a non-native language~\cite{Ha2018FDWBurdenSingapore} restrict opportunities for emotional expression and access to social support.
Caregiving occurs within employment relationships rather than kinship ties, where expectations of professionalism and power asymmetries can further limit safe emotional disclosure~\cite{TAM20181269}.
As a result, FDWs may experience heightened emotional caregiving burden even when the objective demands of care are comparable to those faced by familial caregivers.
Despite these insights, existing caregiving frameworks offer limited guidance on how FDWs access emotionally validating or expressive support in their everyday caregiving practices.




\subsection{Technologies for Caregiving and Emotional Support}
\rev{Prior HCI research has examined technologies that support caregivers’ everyday practices, including role management, experiential knowledge sharing, and the emotional labor of care~\cite{HsuDancingRoles2024, ShinEveryCloud2021, SmritiEmotionWork2024}. Beyond familial caregiving contexts, CSCW scholarship has explored how domestic workers engage with digital technologies under conditions of labor precarity, where adoption is shaped by trust, autonomy, and perceived risk~\cite{Espinosa2025}. Together, this body of work shows that caregiving technologies mediate not only coordination and emotional labor, but also the broader socio-structural constraints shaping how FDWs engage with conversational systems. }



In parallel, advances in LLM–driven conversational agents have explored chatbots as tools for providing empathetic and emotionally responsive interaction~\cite{app14135889}.
Prior studies suggest that chatbots can encourage emotional disclosure and foster a sense of being heard, particularly in contexts where users fear judgment or lack access to human support~\cite{10.1093/joc/jqy026, Ana_Marco_2019}.
However, this work has largely focused on clinical mental health settings or system effectiveness, with limited attention to how caregivers experience and appropriate LLM-driven emotional support in everyday, non-clinical contexts.

For FDWs in particular, it remains unclear how LLM-driven chatbots are experienced as conversational partners under conditions of linguistic constraint, how tolerance for imperfect or fragmented language shapes access to support, and how such systems may be flexibly appropriated to address caregiving-related uncertainty, isolation, and emotional strain.
Addressing these gaps requires attending to FDWs’ lived experiences of interaction with LLM-driven systems, rather than evaluating chatbot performance or outcomes alone.



\section{Methodology}

\subsection{Study Design and Qualitative Data Collection}
We conducted a single-session exploratory qualitative study to examine how FDWs engage with an LLM-driven chatbot, \sophia, when reflecting on emotionally challenging caregiving experiences.  \rev{This format was adopted to minimize demands on participants’ limited free time and to capture initial, situated impressions of the chatbot within an ethically appropriate research setting.} 
Participants were FDWs working in Singapore with experience in eldercare and basic English proficiency.
Seven FDWs (all female), aged 23 to 50, were recruited through in-person intercept recruitment in public spaces frequented by FDWs and via snowball sampling.
These participants were selected as they completed the 12-item short-form Zarit Burden Interview (ZBI-12)~\cite{bedard2001zbi}, with a median score of 14, indicating overall mild caregiver burden.

Each participant took part in one in-person session lasting approximately 1 to 1.5 hours. Sessions included a semi-structured interview guided by the Critical Incident Technique~\cite{doi:10.1177/1468794105056924, flanagan1954cit}.
Based on the recalled caregiving incident, participants then engaged in a guided interaction with \sophia, which was configured to provide empathetic, non-clinical emotional support.
\sophia operated under a single, fixed system prompt that framed it as a non-judgmental conversational partner, encouraged emotional articulation through open-ended questions, and explicitly declined to provide medical, legal, or financial advice. \rev{The restriction was implemented to mitigate ethical risks and to ensure the system remained within a non-clinical, emotionally supportive scope.}
During the interaction, participants were encouraged to verbalize their thoughts using the think-aloud method~\cite{someren1994thinkaloud}.
Sessions concluded with a brief reflective interview about \sophia\textbf{’s} perceived usefulness and limitations.
All sessions were audio-recorded and transcribed, and informed consent was obtained in accordance with IRB-approved procedures. \rev{The research team consisted of university-based researchers trained in HCI and qualitative methods. As we do not share the lived experience of FDWs, we remained attentive to power dynamics during recruitment, interviews, and analysis.}

\subsection{Data Analysis}
We analyzed the data using inductive thematic analysis~\cite{Braun01012006}. 
Audio recordings were transcribed verbatim and iteratively coded to identify patterns related to participants’ caregiving experiences and interactions with \sophia.
Initial codes were compared across participants and refined into higher-level themes through repeated reference to the transcripts to ensure grounding in participants' accounts.
All participants were assigned unique identifiers to preserve anonymity.


\section{Results}
We identified three themes characterizing how FDWs experienced and engaged with \sophia in relation to caregiving burden: (1) psychologically safe and emotionally validating interaction, (2) linguistic accessibility through tolerance of imperfect language, and (3) multifunctional appropriation beyond emotional support.

\subsection{Psychologically Safe and Emotionally Validating Interaction}
Participants commonly experienced \sophia as a psychologically safe and emotionally validating space for reflecting on caregiving challenges. 
Several participants (n=2) described disclosing thoughts and emotions to \sophia that they would not share with others, emphasizing the absence of social judgment or evaluation.
As one participant explained, ``Yes, I can talk to it about anything... I don't want to say to other people but I want to say what is on my mind. So talk with AI is okay, outside other people no one can know better.'' (P1). 

Nearly all participants (n=5) described \sophia's responses as reassuring and stress-reducing, helping them continue caregiving tasks.
Some participants noted that these interactions helped reduce stress and gave them emotional strength to continue caregiving tasks.
Notably, participants emphasized that emotional acknowledgment mattered more than cultural specificity or conversational naturalness, suggesting that perceived validation played a central role in shaping positive experiences.

\subsection{LLM-driven Chatbots Support Linguistic Accessibility by Accepting Imperfect Language}
Linguistic accessibility emerged as a key factor shaping participants' engagement with \sophia.
While some participants (n=3) initially found it challenging to formulate detailed prompts in English, they reported greater comfort interacting in their native languages, which allowed them to express emotions more naturally and with less effort.

At the same time, more than half of the participants (n=4) described being able to communicate effectively in English despite limited proficiency.
Participants frequently used short, fragmented, or grammatically non-standard input, yet perceived \sophia's responses as emotionally supportive and relevant.
Several expressed surprise that \sophia could infer meaning from minimal or imperfect input.
As one participant noted, `Because we're just talking a little bit, but it can reply a long long thing that is correct.'' (P3).
This tolerance for imperfect language reduced pressure to perform linguistic competence and lowered the effort required to seek support. However, linguistic accessibility was not uniform. One participant reported difficulty understanding specific English vocabulary in \sophia's responses, indicating that accessibility depended not only on accommodating user input but also on generating comprehensible responses.

\subsection{LLM Chatbots as Multifunctional Support Resources in Caregiving}
Beyond emotional disclosure, participants appropriated \sophia as a multifunctional support resource, using it to seek reassurance, cope with loneliness, practice English, and navigate caregiving-related uncertainties (n=4).
These uses were not framed as separate from caregiving but as indirectly supporting caregiving by reducing uncertainty, isolation, and emotional strain. 
One participant described relaying on \sophia when other supports were unavailable: ``Yeah I'm happy la, for me it's helping me so much much la...so better to have the app to ask. If I ask the employer, she also don’t know what to do'' (P5).
Participants described turning to \sophia when social, workplace, or informational supports were insufficient, highlighting its role as a flexible and accessible resource within constrained caregiving contexts.




\section{Discussion}

\subsection{Interpreting FDW's Experiences with LLM-Driven Emotional Support}
Our findings suggest that FDWs’ interactions with \sophia supported caregiving burden primarily through expressive emotional support. Participants described feeling heard and validated, aligning with Pearlin’s stress process model~\cite{Pearlin1990StressProcess}, which positions emotional acknowledgment as a key resource when stressors cannot be reduced. For FDWs operating within employment-based care relationships, such support is often constrained by professional norms and power asymmetries. In this context, \sophia functioned as a psychologically safe space for disclosure without social or occupational risk.

Participants valued non-judgmental validation over cultural specificity or conversational naturalness, suggesting that emotional support technologies may be most effective when prioritizing psychological safety over surface-level personalization.
However, this safety is structurally distinct from reciprocal human support. Unlike personal networks, \sophia entails no mutual accountability or relational negotiation. While this asymmetry may lower barriers to disclosure, it reframes emotional support as a one-directional rather than relational exchange.

\subsection{Linguistic Accessibility as a Condition for Emotional Support}
Our findings highlight linguistic accessibility as a key condition enabling emotional support.
Many participants interacted using fragmented or non-standard language, yet still experienced \sophia's responses as emotionally supportive and meaningful.
This tolerance for imperfect input reduced the effort required to articulate distress and lowered barriers to seeking support across language differences.

These findings suggest that linguistic accessibility is not merely a usability concern but is closely tied to emotional access.
For migrant caregivers operating in non-native languages, the ability to communicate without performing linguistic competence can shape whether emotional support feels approachable or burdensome.
At the same time, difficulties understanding certain chatbot responses indicate that accessibility depends on both input tolerance and output clarity.
Designing for linguistic accessibility therefore requires attention to how conversational systems accommodate limited proficiency while generating responses that remain emotionally and cognitively accessible.

\subsection{LLM Chatbots as Flexible Support Resources in Caregiving Contexts}
Participants' appropriated \sophia as a multifunctional resource extending beyond emotional disclosure, using it to seek reassurance, cope with loneliness, and navigate caregiving-related uncertainty when other supports were unavailable. These uses indirectly supported caregiving by addressing isolation and informational gaps.

This flexible appropriation reflects prior work showing that caregiving burden arises from a combination of emotional, relational, and communicative stressors~\cite{Ha2018FDWBurdenSingapore, Pearlin1990StressProcess}.
Rather than replacing existing support networks, \sophia appeared to fill gaps created by linguistic barriers, social hierarchies, and limited access to caregiving knowledge.
These findings suggest that emotional support technologies for caregivers may be most valuable when designed as adaptable, non-judgmental resources that caregivers can appropriate according to their lived constraints and evolving needs.

\subsection{Design Implications for Emotional Support Technologies for FDWs}
Taken together, our findings point to several implications for the design of LLM-driven emotional support tools for FDWs.
First, designers should foreground psychological safety and emotional validation as core interactional properties, enabling caregivers to express distress without fear of judgment or professional failure.
Second, linguistic accessibility should be prioritized over linguistic correctness, accommodating imperfect input while generating clear and emotionally accessible responses.
Finally, emotional support technologies should support flexible, multifunctional appropriation, recognizing that caregiving burden is shaped by broader everyday stressors beyond emotional disclosure alone.
Designing conversational systems as adaptable support resources may better align with FDWs' lived realities and caregiving contexts.
\rev{\subsection{Ethical and Temporal Considerations}
Although participants experienced Sophia as psychologically safe within the session, emotionally responsive systems introduce a form of bounded emotional engagement. Emotional disclosure may extend beyond the temporal limits of a study interaction, raising questions about containment and continuity. Designing such systems therefore requires attention not only to validation and accessibility, but also to how emotional support is introduced, framed, and withdrawn.
While LLM-driven emotional support may provide expressive relief, it does not address the structural labour conditions that give rise to caregiving burden. There is a risk that positioning such systems as solutions may inadvertently individualize responsibility for coping, shifting attention away from broader socio-economic inequities shaping FDWs’ work. Emotional support technologies should therefore be understood as supplementary resources rather than substitutes for structural change.}

\subsection{Limitations and Future Work}
This study has several limitations.
First, participants' interactions with \sophia occurred within a single, time-limited session, capturing initial impressions rather than longer-term patterns of use, trust, or reliance that may emerge over time.
Second, the small sample size and focus on FDWs working in Singapore limit the generalizability of our findings to other caregiving contexts or migrant worker populations.
Finally, participants engaged with a single chatbot configuration, and different prompt designs or system behaviors may shape caregiving experiences in other ways.

Future work could adopt longitudinal field studies to examine how FDWs integrate LLM-driven emotional support tools into everyday caregiving routines over time, and how such systems influence emotional well-being, coping practices, and caregiving burden. Exploring alternative designs grounded in caregiving and self-care frameworks (e.g., Orem’s Self-Care Deficit Nursing Theory~\cite{Orem2001ConceptsOfPractice}) may further inform the development of emotionally supportive technologies for FDWs.


\section{Conclusion}
This paper examined how FDWs in Singapore engage with an LLM-driven chatbot as a form of everyday emotional support in caregiving contexts.
Our findings show that FDWs experienced \sophia as a psychologically safe and emotionally validating space, valued its tolerance for imperfect and non-standard language, and appropriated it as a flexible support resource addressing caregiving-related uncertainty and isolation.
Together, these insights highlight the potential of LLM-driven conversational systems to support caregivers’ emotional well-being under conditions of social, linguistic, and structural constraint.
We conclude by emphasizing the importance of designing emotional support technologies that foreground psychological safety, linguistic accessibility, and flexible appropriation to better align with the lived realities of FDWs and other marginalized caregivers.

\bibliographystyle{ACM-Reference-Format}
\bibliography{references}

@article{Devasahayam22022010,
author = {Theresa W. Devasahayam},
title = {Placement and/or protection? Singapore's labour policies and practices for temporary women migrant workers},
journal = {Journal of the Asia Pacific Economy},
volume = {15},
number = {1},
pages = {45--58},
year = {2010},
publisher = {Routledge},
doi = {10.1080/13547860903488229},


URL = { 
    
        https://doi.org/10.1080/13547860903488229
    
    

},
eprint = { 
    
        https://doi.org/10.1080/13547860903488229
    
    

}

}

@article{Yeoh01052010,
author = {Brenda S. A. Yeoh and Shirlena Huang},
title = {Transnational Domestic Workers and the Negotiation of Mobility and Work Practices in Singapore’s Home‐Spaces},
journal = {Mobilities},
volume = {5},
number = {2},
pages = {219--236},
year = {2010},
publisher = {Routledge},
doi = {10.1080/17450101003665036},


URL = { 
    
        https://doi.org/10.1080/17450101003665036
    
    

},
eprint = { 
    
        https://doi.org/10.1080/17450101003665036
    
    

}

}

@article{TAM20181269,
title = {“I Can't Do This Alone”: a study on foreign domestic workers providing long-term care for frail seniors at home},
journal = {International Psychogeriatrics},
volume = {30},
number = {9},
pages = {1269-1277},
year = {2018},
note = {Issue Theme: Quality of Care for Frail Older Adults},
issn = {1041-6102},
doi = {https://doi.org/10.1017/S1041610217002459},
url = {https://www.sciencedirect.com/science/article/pii/S1041610224017940},
author = {Wai Jia Tam and Gerald Choon-Huat Koh and Helena Legido-Quigley and Ngoc Huong Lien Ha and Philip Lin Kiat Yap}
}

@article{Lawton1989CaregivingAppraisal,
  author  = {Lawton, M. Powell and Kleban, Michael H. and Moss, Miriam and Rovine, Michael and Glicksman, Allen},
  title   = {Measuring caregiving appraisal},
  journal = {Journal of Gerontology},
  volume  = {44},
  number  = {3},
  pages   = {P61--P71},
  year    = {1989},
  month   = {May},
  doi     = {10.1093/geronj/44.3.p61},
}

@article{Pearlin1990StressProcess,
  author  = {Pearlin, Leonard I. and Mullan, Joseph T. and Semple, Shirley J. and Skaff, Marilyn M.},
  title   = {Caregiving and the stress process: an overview of concepts and their measures},
  journal = {The Gerontologist},
  volume  = {30},
  number  = {5},
  pages   = {583--594},
  year    = {1990},
  month   = {October},
  doi     = {10.1093/geront/30.5.583},
}

@article{Zarit1980Burden,
  author  = {Zarit, Steven H. and Reever, Kathleen E. and Bach-Peterson, Joan},
  title   = {Relatives of the impaired elderly: correlates of feelings of burden},
  journal = {The Gerontologist},
  volume  = {20},
  number  = {6},
  pages   = {649--655},
  year    = {1980},
  month   = {December},
  doi     = {10.1093/geront/20.6.649},
}

@article{Ha2018FDWBurdenSingapore,
  author  = {Ha, Ngoc Huong Lien and Chong, Mei Sian and Choo, Robin Wai Munn and Tam, Wai Jia and Yap, Philip Lin Kiat},
  title   = {Caregiving burden in foreign domestic workers caring for frail older adults in Singapore},
  journal = {International Psychogeriatrics},
  volume  = {30},
  number  = {8},
  pages   = {1139--1147},
  year    = {2018},
  month   = {August},
  doi     = {10.1017/S1041610218000200},
}

@Article{app14135889,
AUTHOR = {Casu, Mirko and Triscari, Sergio and Battiato, Sebastiano and Guarnera, Luca and Caponnetto, Pasquale},
TITLE = {AI Chatbots for Mental Health: A Scoping Review of Effectiveness, Feasibility, and Applications},
JOURNAL = {Applied Sciences},
VOLUME = {14},
YEAR = {2024},
NUMBER = {13},
ARTICLE-NUMBER = {5889},
URL = {https://www.mdpi.com/2076-3417/14/13/5889},
ISSN = {2076-3417},
DOI = {10.3390/app14135889}
}

@Article{wysa2018,
author="Inkster, Becky
and Sarda, Shubhankar
and Subramanian, Vinod",
title="An Empathy-Driven, Conversational Artificial Intelligence Agent (Wysa) for Digital Mental Well-Being: Real-World Data Evaluation Mixed-Methods Study",
journal="JMIR Mhealth Uhealth",
year="2018",
month="Nov",
day="23",
volume="6",
number="11",
pages="e12106",
issn="2291-5222",
doi="10.2196/12106",
url="http://mhealth.jmir.org/2018/11/e12106/",
url="https://doi.org/10.2196/12106",
url="http://www.ncbi.nlm.nih.gov/pubmed/30470676"
}

@article{Ho2022PeerSupportMDW,
  author  = {Ho, Ken Hok Man and Yang, Chen and Leung, Alex Kwun Yat and Bressington, Daniel and Chien, Wai Tong and Cheng, Qijin and Cheung, Daphne Sze Ki},
  title   = {Peer Support and Mental Health of Migrant Domestic Workers: A Scoping Review},
  journal = {International Journal of Environmental Research and Public Health},
  volume  = {19},
  number  = {13},
  pages   = {7617},
  year    = {2022},
  month   = {June},
  doi     = {10.3390/ijerph19137617},
}

@article{Shrestha2019WorkerMobilitySingapore,
  author  = {Shrestha, Slesh A. and Yang, Dean},
  title   = {Facilitating Worker Mobility: A Randomized Information Intervention among Migrant Workers in Singapore},
  journal = {Economic Development and Cultural Change},
  volume  = {68},
  number  = {1},
  pages   = {63--91},
  year    = {2019},
  month   = {October},
  doi     = {10.1086/700620},
}

@article{bedard2001zbi,
  author  = {B{\'e}dard, Martin and Molloy, David W. and Squire, Larry and Dubois, Sylvie and Lever, Jeffrey A. and O'Donnell, Mark},
  title   = {The Zarit Burden Interview: A New Short Version and Screening Version},
  journal = {The Gerontologist},
  volume  = {41},
  number  = {5},
  pages   = {652--657},
  year    = {2001},
  month   = oct,
  doi     = {10.1093/geront/41.5.652}
}

@article{doi:10.1177/1468794105056924,
author = {Lee D. Butterfield and William A. Borgen and Norman E. Amundson and Asa-Sophia T. Maglio},
title ={Fifty years of the critical incident technique: 1954-2004 and beyond},

journal = {Qualitative Research},
volume = {5},
number = {4},
pages = {475-497},
year = {2005},
doi = {10.1177/1468794105056924},

URL = { 
    
        https://doi.org/10.1177/1468794105056924
    
    

},
eprint = { 
    
        https://doi.org/10.1177/1468794105056924
    
    

}
}

@article{flanagan1954cit,
  author  = {Flanagan, John C.},
  title   = {The Critical Incident Technique},
  journal = {Psychological Bulletin},
  volume  = {51},
  number  = {4},
  pages   = {327--358},
  year    = {1954},
  doi     = {10.1037/h0061470}
}

@book{someren1994thinkaloud,
author = {Someren, Maarten and Barnard, Yvonne and Sandberg, Jacobijn},
year = {1994},
month = {01},
pages = {},
title = {The Think Aloud Method - A Practical Guide to Modelling CognitiveProcesses}
}

@article{Parreñas16122021,
author = {Rhacel Salazar Parreñas and Krittiya Kantachote and Rachel Silvey},
title = {Soft violence: migrant domestic worker precarity and the management of unfree labour in Singapore},
journal = {Journal of Ethnic and Migration Studies},
volume = {47},
number = {20},
pages = {4671--4687},
year = {2021},
publisher = {Routledge},
doi = {10.1080/1369183X.2020.1732614},


URL = { 
    
        https://doi.org/10.1080/1369183X.2020.1732614
    
    

},
eprint = { 
    
        https://doi.org/10.1080/1369183X.2020.1732614
    
    

}

}

@article{10.1093/joc/jqy026,
    author = {Ho, Annabell and Hancock, Jeff and Miner, Adam S},
    title = {Psychological, Relational, and Emotional Effects of Self-Disclosure After Conversations With a Chatbot},
    journal = {Journal of Communication},
    volume = {68},
    number = {4},
    pages = {712-733},
    year = {2018},
    month = {05},
    issn = {0021-9916},
    doi = {10.1093/joc/jqy026},
    url = {https://doi.org/10.1093/joc/jqy026},
    eprint = {https://academic.oup.com/joc/article-pdf/68/4/712/25410448/jqy026.pdf},
}

@article{Ana_Marco_2019,
  author       = {Ana Paula Chaves and
                  Marco Aur{\'{e}}lio Gerosa},
  title        = {How should my chatbot interact? {A} survey on human-chatbot interaction
                  design},
  journal      = {CoRR},
  volume       = {abs/1904.02743},
  year         = {2019},
  url          = {http://arxiv.org/abs/1904.02743},
  eprinttype    = {arXiv},
  eprint       = {1904.02743},
  bibsource    = {dblp computer science bibliography, https://dblp.org}
}

@article{SmritiEmotionWork2024,
author = {Smriti, Diva and Wang, Lu and Huh-Yoo, Jina},
title = {Emotion Work in Caregiving: The Role of Technology to Support Informal Caregivers of Persons Living With Dementia},
year = {2024},
issue_date = {April 2024},
publisher = {Association for Computing Machinery},
address = {New York, NY, USA},
volume = {8},
number = {CSCW1},
url = {https://doi.org/10.1145/3637325},
doi = {10.1145/3637325},
journal = {Proc. ACM Hum.-Comput. Interact.},
month = apr,
articleno = {48},
numpages = {34}
}

@inproceedings{HsuDancingRoles2024,
author = {Hsu, Long-Jing and Chung, Chia-Fang},
title = {Dancing with the Roles: Towards Designing Technology that Supports the Multifaceted Roles of Caregivers for Older Adults},
year = {2024},
isbn = {9798400703300},
publisher = {Association for Computing Machinery},
address = {New York, NY, USA},
url = {https://doi.org/10.1145/3613904.3642728},
doi = {10.1145/3613904.3642728},
booktitle = {Proceedings of the 2024 CHI Conference on Human Factors in Computing Systems},
articleno = {1010},
numpages = {12},
location = {Honolulu, HI, USA},
series = {CHI '24}
}

@article{ShinEveryCloud2021,
author = {Shin, Ji Youn and Chaar, Dima and Davis, Catherine and Choi, Sung Won and Lee, Hee Rin},
title = {Every Cloud Has a Silver Lining: Exploring Experiential Knowledge and Assets of Family Caregivers},
year = {2021},
issue_date = {October 2021},
publisher = {Association for Computing Machinery},
address = {New York, NY, USA},
volume = {5},
number = {CSCW2},
url = {https://doi.org/10.1145/3479560},
doi = {10.1145/3479560},
journal = {Proc. ACM Hum.-Comput. Interact.},
month = oct,
articleno = {416},
numpages = {25}
}

@Article{woebott2017,
author="Fitzpatrick, Kathleen Kara
and Darcy, Alison
and Vierhile, Molly",
title="Delivering Cognitive Behavior Therapy to Young Adults With Symptoms of Depression and Anxiety Using a Fully Automated Conversational Agent (Woebot): A Randomized Controlled Trial",
journal="JMIR Ment Health",
year="2017",
month="Jun",
day="06",
volume="4",
number="2",
pages="e19",
issn="2368-7959",
doi="10.2196/mental.7785",
url="http://mental.jmir.org/2017/2/e19/",
url="https://doi.org/10.2196/mental.7785",
url="http://www.ncbi.nlm.nih.gov/pubmed/28588005"
}

@article{Braun01012006,
author = {Virginia Braun and Victoria Clarke},
title = {Using thematic analysis in psychology},
journal = {Qualitative Research in Psychology},
volume = {3},
number = {2},
pages = {77--101},
year = {2006},
publisher = {Routledge},
doi = {10.1191/1478088706qp063oa},


URL = { 
    
        https://doi.org/10.1191/1478088706qp063oa
    
    

},
eprint = { 
    
        https://doi.org/10.1191/1478088706qp063oa
    
    

}

}

@book{Orem2001ConceptsOfPractice,
  author    = {Orem, Dorothea E.},
  title     = {Nursing: Concepts of Practice},
  edition   = {6},
  year      = {2001},
  publisher = {Mosby},
  address   = {St. Louis, MO}
}

@article{Espinosa2025,
author = {Fernandez-Espinosa, Mariana and Gonzalez-Bejar, Mariana and Wiesner, Jacobo and G\'{o}mez-Zar\'{a}, Diego},
title = {When Technologies Are Not Enough: Understanding How Domestic Workers Employ (and Avoid) Online Technologies in Their Work Practices},
year = {2025},
issue_date = {November 2025},
publisher = {Association for Computing Machinery},
address = {New York, NY, USA},
volume = {9},
number = {7},
url = {https://doi.org/10.1145/3757523},
doi = {10.1145/3757523},
journal = {Proc. ACM Hum.-Comput. Interact.},
month = oct,
articleno = {CSCW342},
numpages = {34}
}







\end{document}